\documentclass[preprint]{vgtc}               

\graphicspath{{figures/}{pictures/}{images/}{./}} 

\usepackage{times}                     

\usepackage{tabu}                      
\usepackage{booktabs}                  
\usepackage{lipsum}                    
\usepackage{mwe}                       

\usepackage{mathptmx}                  

\usepackage{subcaption}

\onlineid{1131}

\vgtccategory{Research}

\vgtcinsertpkg

\preprinttext{To appear in the proceedings of the IEEE International Symposium on Mixed and Augmented Reality (ISMAR) 2025}

\hypersetup{
    colorlinks=false,
    pdfborder={0 0 0},
    hidelinks
}

\title{Sonify Anything: {Towards Context-Aware Sonic Interactions in AR}}

\author{Laura Schütz\thanks{e-mail: laura.schuetz@tum.de}
\and Sasan Matinfar
\and Ulrich Eck
\and Daniel Roth
\and Nassir Navab}
\affiliation{\scriptsize Technical University of Munich}

\teaser{
  \centering
  \includegraphics[width=\linewidth]{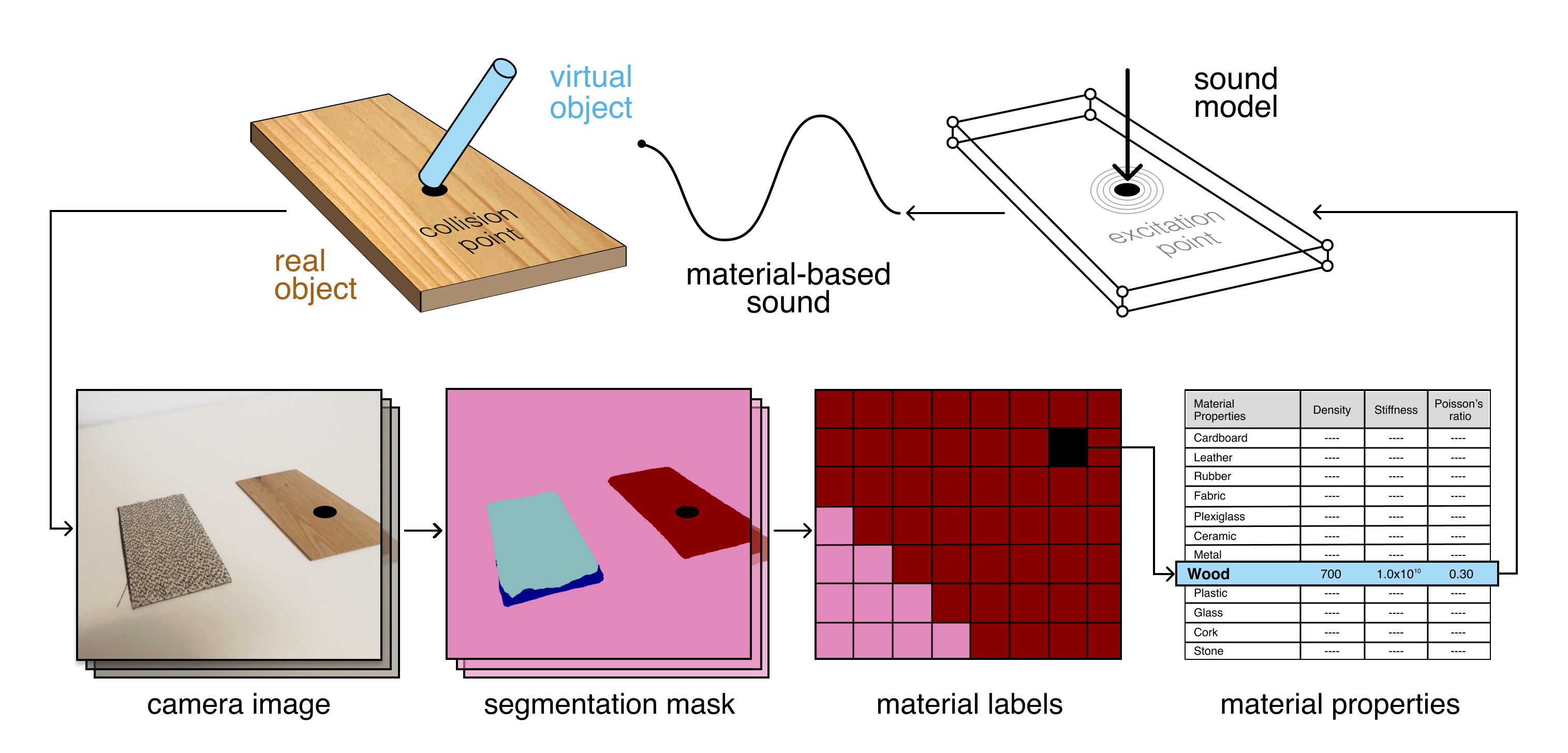}
  \caption{Material-based sonification approach for real-virtual object interactions in Augmented Reality}
  \label{fig:teaser}
}

\abstract{
In Augmented Reality (AR), virtual objects interact with real objects. However, the lack of physicality of virtual objects leads to the absence of natural sonic interactions. When virtual and real objects collide, either no sound or a generic sound is played. Both lead to an incongruent multisensory experience, reducing interaction and object realism. Unlike in Virtual Reality (VR) and games, where predefined scenes and interactions allow for the playback of pre-recorded sound samples, AR requires real-time sound synthesis that dynamically adapts to novel contexts and objects to provide audiovisual congruence during interaction. To enhance real-virtual object interactions in AR, we propose a framework for context-aware sounds using methods from computer vision to recognize and segment the materials of real objects. The material's physical properties and the impact dynamics of the interaction are used to generate material-based sounds in real-time using physical modelling synthesis. In a user study with 24 participants, we compared our congruent material-based sounds to {a} generic sound effect{, mirroring the current standard of non-context-aware sounds in AR applications}. The results showed that {material-based} sounds led to significantly more realistic sonic interactions. {Material-based} sounds also enabled participants to distinguish visually similar materials with significantly greater accuracy and confidence.
These findings show that context-aware, material-based sonic interactions in AR foster a stronger sense of realism and enhance our perception of real-world surroundings.
}


\begin{document}


\firstsection{Introduction}
\label{sec:intro}

\maketitle


Our interactions with physical objects naturally create sound. 
Sound occurs when vibrations set a medium like air into motion. The physical properties of a medium directly shape the sound it produces. For instance, placing a cup on a table, typing on a keyboard, or knocking on a wooden door all make distinct sounds. Therefore, sounds contain information about the source and the event that produced it, enabling people to identify materials from sounds. Audition helps us pick up these sounds and understand what’s happening in our environment. The idea of ecological acoustics explains how we use sound to make sense of the world \cite{McDermott2024ecologicalacoustics}. It explores how sounds carry information about objects, spaces, and events, and how our auditory system interprets them to navigate the world.

Materials are mainly identified in impact sounds via frequency and damping parameters of the sound \cite{klatzky2000perception, McAdams2004psychomechanics}. However, since frequency changes can also be attributed to changes in object geometry, damping cues have been shown to be more reliable for material identification from hitting actions \cite{mcadams2010psychomechanics}. It has furthermore been shown that recovering materials from action sounds is harder than identifying the sound producing action, like scraping or rolling \cite{lemaitre2012auditory}.

Augmented Reality (AR) applications aim to seamlessly blend virtual and physical elements. However, when interacting with virtual objects, we are robbed of these ecological sounds. The relationship between the sound and the sound source is lost when we use no sound or incongruent sound effects to sonify virtual object interactions, resulting in incongruent multisensory perception. This discrepancy reduces the interaction realism and weakens the user's sense of presence.

Studies have demonstrated that audiovisual congruence increases presence and realism in Virtual Reality (VR) experiences \cite{kern2020audio}. Semantic congruence specifically has been shown to improve recognition speed and attentional control, making it a key factor in enhancing interaction realism \cite{laurienti2004semantic, chen2010hearing}.
Unlike in VR and games, where objects and environments are predefined, and sound samples can be retrieved from a sound library when a sound-producing event occurs, AR operates in constantly changing real-world environments. As a result, pre-recorded audio samples are insufficient to create realistic sonic interactions. There is a need for real-time audio synthesis that dynamically considers the materiality of real-world objects during interactions in AR.

This work aims to bring audiovisual congruence to AR interactions and thereby enhance the perceptual link between users and their physical environment.
We introduce a material-based sonification technique that generates dynamic, congruent impact sounds for real-virtual object interactions in AR, using object material properties and impact dynamics to produce accurate audio feedback. Leveraging existing machine learning methods for real-time material segmentation, we extract material information from unseen environments. By combining these techniques with physics-based sound synthesis, we enable real-time, material-driven audio feedback without relying on pre-recorded audio clips or large training datasets.
{What we do not contribute is a technique for generating highly realistic material sounds, nor a comprehensive study on the perceptual effects of audiovisual congruence in AR.
Instead, o}ur work makes the following contributions:
\begin{enumerate}
\item We propose a framework for context-aware sounds in AR using material segmentation and physical modelling sound synthesis.
\item We demonstrate a material-based sonification approach that creates physics-based sounds in real-time.
\item We report results from a user study {showing that the proposed material-based sounds improve perceived interaction realism over standard generic sounds.}

\end{enumerate}

\section{Related Work}
\label{sec:related_work}

\subsection{Material Segmentation}
Material segmentation is a field of study in computer vision that has seen significant advancements in recent years. It is concerned with the task of assigning material labels (e.g., wood, metal) to each pixel in an image. Material segmentation is, for example, used in robotics, for effective decision-making and object interaction \cite{ravipati2024robots}, or in autonomous driving, where material cues help improve terrain understanding and safety decisions \cite{cai2024cars}.
The Materials in Context Database (MINC) is a large dataset that facilitates deep learning approaches for material recognition \cite{bell2015material}.  
Another recent contribution is the Dense Material Segmentation dataset, which provides 3.2 million dense material annotations for a diverse set of indoor and outdoor scenes, objects, viewpoints, and materials \cite{upchurch2022dms}. This makes it especially suitable for augmented reality use cases, where detailed material recognition from diverse viewing angles is required for realistic interactions.

\subsection{Sound Synthesis Techniques}
\label{sec:sound_synth}

To recreate realistic action sounds, we can make use of a variety of sound synthesis methods. 
Three prominent sound synthesis methods for action sounds are:

\textbf{Sampling-Based Sound Synthesis:} 
Widely used in games and virtual reality, this method is an easy way to create action sounds from pre-recorded samples \cite{lloyd2011sound}. The samples are associated with events or locations in the scene and played back when the interaction occurs. To simulate continuous sounds, periodic elements of the waveform are often looped. Filters and envelopes can be applied to diversify the sound output from a given set of samples \cite{cook2002sound}.
Although sample-based synthesis is easy to implement and computationally efficient, it is limited by the prerecorded sound clips available in a database. 
As a result, it lacks the flexibility to respond to unexpected changes in the environment, making it less suitable for context-sensitive applications such as Augmented Reality. 

\textbf{Data-Driven Sound Synthesis:}
This approach to sound synthesis uses various data analysis techniques, including statistical methods and machine learning, to infer action sounds. Physics-driven machine learning approaches, such as physics-informed diffusion models, have been developed to synthesize impact sounds from videos \cite{kun2023physicsdiff}. Identifying visual representations of sound-producing actions can be learned from egocentric videos \cite{chen2024soundingactions} and used to generate action-matching sounds \cite{chen2024action2sound}. 
Large language models (LLMs) have been employed to query for {foley} sound effects matching the content of a video clip. These sound samples are later adjusted to match the motion dynamics in the video clip \cite{lin2025foley, chen2025video}.
While data-driven approaches can create realistic action sounds from video data, they are {constrained by dataset limitations and inference latency}.

\textbf{Physical Modelling Sound Synthesis:} 
Given that material sounds are closely tied to the physicality of the sound-producing objects and actions, model-based synthesis is a promising approach to creating material-based action sounds. 
This method can create highly realistic impact sounds in real-time, but is more computationally intensive than sample-based sonification. However, several techniques for accelerating physically based sound simulation have been shown to reduce the computational cost, enabling simultaneous simulation of numerous sound models \cite{Raghuvanshi2007physically}, without relying on prior training, as is necessary for data-driven techniques. 

The construction of sound models is commonly achieved through physical modelling synthesis \cite{cook1996physically}. This technique aims to mimic the physical characteristics of real-world instruments, effectively emulating the behavior of actual objects. 
Several simulation software have been proposed for modeling sounds dynamically based on object properties. Early work in modal synthesis, such as Mosaic \cite{morrison1993mosaic} and Modalys \cite{eckel1995sound}, demonstrated its potential for realistic audio generation. Modalys, in particular, uses the finite element method to perform physical modelling synthesis. By solving differential equations associated with vibrating systems, it can represent key dynamic characteristics such as natural frequencies, damping behavior, and mode shapes, relevant to the creation of realistic material sounds. By precomputing the object's modes of frequencies, this approach enables efficient synthesis, ideal for realtime interactions.

Van den Doel et al. \cite{doel2001foley} introduced a method for real-time synthesis of contact sounds, such as impact, rolling, and friction, for solid materials using modal synthesis. Later work on modal synthesis for interactive sounds investigated the inclusion of surface information at three levels of resolution (object shape, visible surface bumpiness, microscopic roughness) for synthesizing complex contact sounds in virtual environments \cite{ren2010virtual}. More recently, learning-based methods have been proposed for real-time modal impact sound synthesis \cite{jin2020deep}. 

However, a key challenge in using modal techniques is the lack of automatic determination of satisfactory material parameters that recreate realistic audio of sound-producing materials \cite{serafin2018sonic}. In AR, this problem is exacerbated. Unlike virtual environments where predefined 3D objects are assigned parameters that will lead to satisfactory audio output, we deal with unknown physical objects, interactions, and collision points in AR. Therefore, we believe that a real-time understanding of the context, objects, and interactions is needed to obtain the required information to create context-aware sonic interactions.

{While the works referenced in this section from the fields of computer vision \cite{kun2023physicsdiff, chen2024action2sound, chen2024soundingactions, chen2025video}, computer graphics \cite{cook2002sound, Raghuvanshi2007physically, doel2001foley}, and computer music \cite{cook1996physically, morrison1993mosaic, eckel1995sound} focus on methods for generating highly-realistic synthesized action sounds, our work, proposing a novel material-based sonic interaction technique, leverages an established sound synthesis technique that has been proven to create high-quality impact sounds \cite{eckel1995sound}.}

\subsection{Audio Interactions in XR}

\textbf{Virtual Reality:} Numerous studies have explored the simulation of realistic sonic interactions in virtual environments. Serafin et al. \cite{serafin2018sonic} stated that an immersive sonic experience relies on action sounds, binaural rendering, environmental sounds, and sound propagation. Since this study focuses on action sounds - sounds produced by the listener and changing with movement \cite{Geronazzo2023SIVE}) - we will highlight a few works on action sounds in this section.

Besides object-related contact sounds, the simulation of footstep sounds in VR has also been studied. Nordahl et al. \cite{nordahl2011walking} presented an algorithm for simulating walking sounds on solid and aggregated surfaces. In addition, the use of velocity-based variations in walking sounds has been proposed for simulated sneaking in VR \cite{cmentowski2021sneaking}.
Schütz et al. \cite{schuetz2024mmii} introduced a multisensory interaction framework for audiovisual interaction with anatomical structures using a physically based sonification approach. 
Their framework, along with the other physics-based approaches for impact sounds outlined in \cref{sec:sound_synth} \cite{doel2001foley, ren2010virtual, jin2020deep} are effective for fully virtual environments. However, they would require high-resolution 3D reconstructions of physical objects in the scene and real-time computation of the objects' natural frequencies to be feasible for application in AR, limiting their practicality.
All the above studies were conducted in entirely simulated or virtual environments. In contrast, only a few works have investigated context-based sounds in AR. 

\textbf{Augmented Reality:} To create context-aware sounds in AR, we require knowledge about the real-world environment. This can be achieved by using sensing technology and analyzing the sensor or camera data.
Wilson et al. \cite{wilson2007swan} were among the first to propose a system that provides relevant audio feedback about the environment. Using GPS and head orientation to estimate the user's pose, the system dynamically generates spatialized, non-speech audio cues that provide blind or visually impaired users with navigation and inform them about nearby features (e.g., benches or stairs).
Medical augmented reality systems using electromagnetic or visual tracking demonstrated precise localization of points of interest within the patient body using parameter mapping sonification \cite{schuetz2024shape, schuetz2023audiovis}.
A study by Su et al. \cite{su2024sonifyAR} introduced a system that generates context-aware sound effects for AR by analyzing the semantics of the virtual augmentations and real-world context. An LLM processes this information to acquire suitable audio through sample retrieval, text-to-sound generation, or text-based sound style transfer.
While their approach targets the curation of a set of sound effects that people can choose from to sonify animated AR content, not requiring real-time sound synthesis, our work focuses on generating real-time impact sounds resulting from human-object interactions. 

To the best of our knowledge, this is the first work to introduce a {context-aware} framework for {material-based} impact sounds in AR. We sonify real-virtual object interactions using real-time material segmentation of physical objects, enabling physically-based sound synthesis in AR. Using this system, semantically and temporally congruent interaction sounds can be generated for unseen environments in real-time.

\subsection{Multisensory Congruence in XR}

Laurienti et al. \cite{laurienti2004semantic} report that semantically congruent audiovisual stimuli enhance recognition speed and accuracy in perception tasks.
Chen and Spence \cite{chen2010hearing} showed that semantically congruent audiovisual stimuli enhance, whereas semantically incongruent audiovisual stimuli impair, object identification performance. They further highlight the role of temporal congruence in audiovisual perception, showing that users can tolerate slight delays of up to 300 ms between audio and visual stimuli, but that excessive desynchronization disrupts the formation of a coherent multisensory percept.
Besides identification accuracy and speed, matching cross-modal stimuli have been shown to enhance the ability to select and hold attention on an object when multiple sensory stimuli compete for attention \cite{van2009multisensory}.
Additionally, a study by Fujisaki et al. \cite{fujisaki2014audiovisual} on audiovisual integration in material perception revealed that sound accuracy significantly affects the perceived material properties of objects.

In VR environments specifically, establishing multisensory congruence has been shown to improve the user's attention and sense of presence \cite{serafin2018sonic}.
A study by Kim et al. \cite{kim2022congruency} demonstrated that audiovisual congruence significantly improved users’ sense of presence, realism, and emotional engagement, while incongruence reduced immersion and increased perceived effort during interaction.
Two studies on gait-aware auditory feedback showed that congruent audio feedback enhances presence and immersion in virtual environments~\cite{hoppe2019VRsneaky, ibers2023effects}. 
These findings underscore the importance of ensuring multisensory congruence in XR environments, as congruent stimuli enhance realism and presence.

{Many of the works from cognitive psychology and neuroscience cited in this section \cite{laurienti2004semantic, chen2010hearing, fujisaki2014audiovisual, van2009multisensory} focus on evaluating the effects of varying degrees of multisensory congruence on user perception in controlled psychological experiments. In contrast, our work compares two audiovisual interaction techniques in AR to demonstrate that congruent, material-based sounds can enhance the perceived realism of AR interactions compared to generic sounds.}

\section{Methods}
\label{sec:methods}

We propose a material-based approach to sonic interactions in AR using material segmentation to inform the physical modelling sound synthesis. \cref{fig:teaser} depicts an overview of the system components. Each component is described in more detail below.

\begin{figure*}[ht!]
 \centering
 \includegraphics[width=\linewidth]{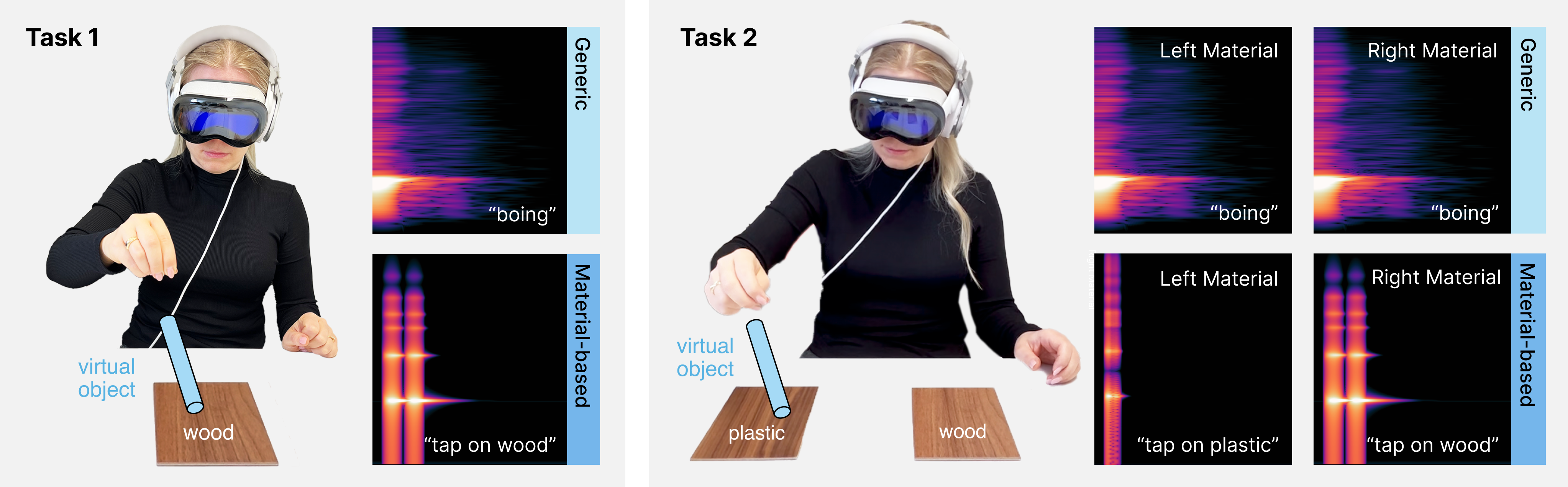}
 \caption{Task 1 (left) and Task 2 (right) of the user study. Task 1 - Participants had to identify one material at a time using the {material-based} and {generic} sound condition. Task 2 - Participants had to distinguish two visually similar materials in the {material-based} and {generic} sound condition. The same sound effect was used for every material in the {generic} condition. Individual {physical modelling}-based sounds were generated for every material in the {material-based} condition.}
 \label{fig:tasks}
\end{figure*}

\subsection{Scene Understanding}
To obtain material information about the objects in the environment, we stream camera images from the left camera of the VisionPro via a WebSocket to a Python script running on a MacBookPro (M1 Max). Camera frames were sent every 200 milliseconds. The Python script runs a pre-trained material segmentation model\footnote{https://github.com/apple/ml-dms-dataset}, presented in a paper by Upchurch and Niu \cite{upchurch2022dms} on the RGB camera images. The frames are originally captured at 1920×1080 resolution and downscaled to 960×540 to enable faster inference. The resulting segmentation mask - an image with color-coded material labels - is then sent back to the Swift script running on the Vision Pro. Lower image resolutions such as 512×512 or 256×256 significantly degraded classification accuracy. Through empirical testing, 960×540 proved to be the most optimal trade-off between processing speed and segmentation quality for our application. 

\subsection{Object Interaction}
The AR application was developed using Unity\footnote{https://unity.com} (v 6000.0.27f1), with Unity PolySpatial (v 2.1.2) supporting deployment on visionOS (v 2.0). Hand interactions were implemented via the XR Interaction Toolkit (v 3.0.7). To receive segmentation masks within Unity, a dedicated package for Apple Vision Pro camera access\footnote{https://github.com/styly-dev/EnterpriseCameraAccessPlugin} was integrated. The package facilitates communication between Swift and Unity through a callback mechanism. The callback function is implemented in C++ (Unity) and invoked in Swift to acquire the segmentation mask images from the Vision Pro. The segmentation masks were received and stored as Texture2D objects. The ARFoundation (v 6.0.5) AR Plane Manager was used for plane detection. When a virtual object collided with a real-world surface (AR plane), the world-space collision point was transformed into the image plane of the previously recorded segmentation masks using the headset camera intrinsics and the corresponding camera-to-world transformation matrix. For each segmentation mask in the history buffer of up to 5 images, the 3×3 neighborhood around the projected UV coordinate of the collision was sampled to retrieve the most frequent RGB pixel color values. These values were then used to retrieve the associated material class from a dictionary mapping RGB values to material names.
A majority vote across all segmentation masks determined the most likely material at the collision point to provide temporal smoothing and robustness against noisy classification results.

\subsection{Material-based Sonification} 
The collision material name is sent from Unity to Max/MSP\footnote{https://cycling74.com/products/max}, a software for audio programming, using Open Sound Control \cite{wright2005}, a network protocol for interactive computer music. Modalys for Max\footnote{https://support.ircam.fr/docs/Modalys/current/}, a physical modelling sound synthesis software, is used to sonify the object interactions. Depending on the object shape, sound models can be created. However, in our study we only focused on evaluating material samples in the shape of plates. Therefore, we used a 3D rectangular plate model in Modalys to represent the geometry of the real-world material samples. 
The physical material properties of density (kg/m3), stiffness (Young modulus, kPa), and Poisson's ratio were used to parameterize the plate model (see \cref{tab:phys_values})). Additional parameters included plate thickness and damping coefficients such as constant loss and frequency loss.
Modalys numerically solves a set of differential equations that control the dynamics of these modes over time, simulating how the plate responds to the excitation based on its physical parameters. 
When a collision occurs in Unity, force is applied to excite the virtual plate object in Modalys. All kinds of interactions, like bowing or striking, can be realized within Modalys.
To isolate the perception of material properties, we used a one-dimensional force connector, a point-mass excitation at a single location on the rectangular plate. In physical modeling synthesis, complex excitations can introduce nonlinearities that mask the acoustic characteristics of the simulated material. By simplifying the excitation, we reduced its impact on the resulting sound, ensuring that sound differences originated mostly from material properties rather than excitation artifacts.
When an excitation occurs, energy is introduced into the model, triggering vibrations across the plate's resonant modes. Modalys picks up the resulting motion at defined listening points, where the sum of the active modes is converted into an audio signal in Max/MSP.

\section{Study}
\label{sec:study}

To determine whether audiovisual congruence established using our material-based sonification approach can enhance interaction realism in AR, we performed a within-subject study comparing {our context-ware, material-based sounds to a standard generic sound} to investigate the following hypotheses:

\begin{itemize}
\item[\textbf{\textit{H1}}] {Material-based sounds} enhance material identification accuracy (Task 1)
\vspace{-0.7em}
\item[\textbf{\textit{H2}}] {Material-based sounds} enhance material identification confidence (Task 1 \& Task 2)
\vspace{-0.7em}
\item[\textbf{\textit{H3}}] {Material-based sounds} enhance sonic interaction realism (Task 1)
\vspace{-0.7em}
\item[\textbf{\textit{H4}}] {Material-based sounds} facilitate distinguishability of visually similar materials (Task 2)
\end{itemize}

In the AR application, participants were given a virtual stick, which they were instructed to grab with a pinch gesture and use to tap on real, physical material samples. Participants were told that they could imagine the stick as a stiff object similar to a plastic pen. The material of the virtual stick was purposefully left undefined as our approach aimed to exclusively explore the simulation of impact sounds on real-world objects.
Participants had to perform two tasks (\cref{fig:tasks}). Both tasks included two audio conditions: {material-based} and {generic}.

\subsection{Task 1}
In Task 1, each condition included 10 trials, one for every material. Participants were presented with one material at a time and asked to tap on the material sample using the virtual stick to create the impact sounds. Based on the visual appearance of the material and the tapping sound, they were asked to answer three questions related to the material and its properties and one question about the realism of the sonic interaction.
After either condition, the participants completed a post-{condition} questionnaire, which included questions related to their material perception confidence and the helpfulness of the sound (see \cref{sec:dependent_vars_t1} for all questions in Task 1).

\subsubsection{Stimulus}
\label{sec:stimulus_t1}
\begin{figure*}[ht!]
 \centering
 \includegraphics[width=\linewidth]{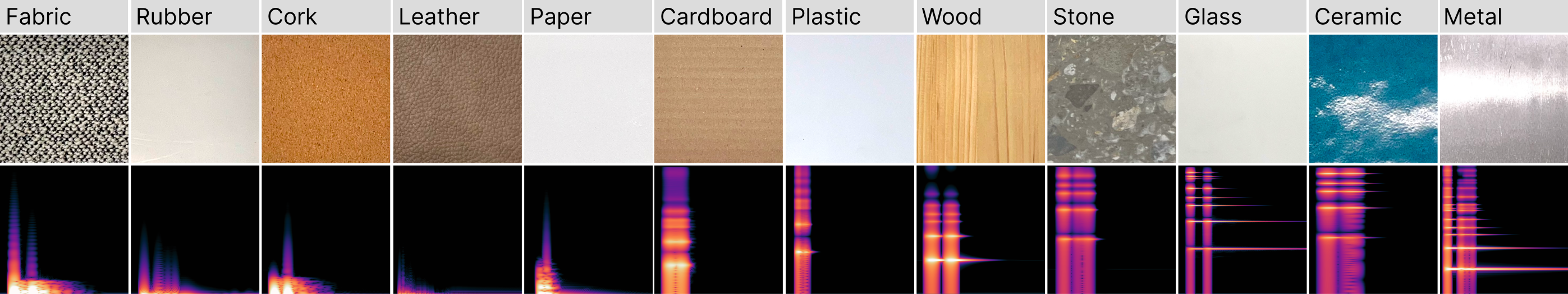}
 \caption{Materials used in Task 1 and Task 2 of the user study along with spectrograms of the  {material-based} sounds generated using physical modelling synthesis. Materials are sorted by increasing stiffness from left to right.}
 \label{fig:materials}
\end{figure*}

Ten realistic indoor surface materials covering a wide range of physical properties, from elastic to stiff, from airy to dense, were included in Task 1: Cardboard, Ceramic, Cork, Fabric, Glass, Leather, Metal, Plastic, Stone, Wood (see \cref{tab:phys_values} for their physical properties and \cref{fig:materials} for images and sounds). The samples were sized 10x10, 11x15, or 22x22. The thickness of the plates ranged from 0.3 to 1.0 cm.

\begin{table}[tb]
  \caption{Physical properties of the materials used in Task 1 and Task 2 based on research in mechanics and material sciences \cite{morey1954properties, paquin1995properties, siegesmund2010physical, niemz2023physical, landel1993mechanical}: Density (kg/m3), Stiffness (Young modulus
(N/m2)), Poisson’s ratio}
  \label{tab:phys_values}
  \scriptsize%
	\centering%
  \begin{tabu}{%
	l%
	*{3}{c}%
	}
  \toprule
   Material & Density (kg/m3) &   Stiffness (N/m2) &   Poisson's ratio   \\
  \midrule
	Cardboard & 689 &  5.0 x 10\textsuperscript{8} & 0.33 \\
  Ceramic & 2600 & 2.0 x 10\textsuperscript{11} & 0.25\\ 
  Cork & 240 & 1.0 x 10\textsuperscript{8}  & 0.30 \\
  Fabric & 1500 & 1.0 x 10\textsuperscript{6} & 0.30\\
  Glass & 2500 & 7.2 x 10\textsuperscript{10} & 0.20\\ 
  Leather & 860 & 1.0 x 10\textsuperscript{8} & 0.40\\
  Metal & 7800 & 2.0 x 10\textsuperscript{11}  & 0.30\\
  Paper & 800 & 5.0 x 10\textsuperscript{8} & 0.33 \\
  Plastic & 1100 &  2.5 x 10\textsuperscript{9} & 0.35\\
  Rubber & 1100& 1.0 x 10\textsuperscript{7}& 0.50\\
  Stone & 2700 &  5.0 x 10\textsuperscript{10} & 0.25\\
  Wood & 700 &  1.0 x 10\textsuperscript{10} & 0.30\\
  \bottomrule
  \end{tabu}%
\end{table}

\subsubsection{Independent Variables}
\label{sec:independent_vars_t1}

\noindent\textbf{{Sound}} The task included two audio conditions: {material-based} and {generic}.
The {material-based} condition used the proposed material-based sonification approach to create congruent material interaction sounds for all twelve materials used in the study.
In the {generic} condition, the same audio sample ("Button Pop" from the XR Interaction Toolkit (v 3.0.7)) was played for all materials, emulating the current standard in AR where identical sounds are used for interactions between virtual and physical objects {regardless of their materiality}.

\subsubsection{Dependent Variables}
\label{sec:dependent_vars_t1}

\noindent\textbf{Material \& Properties} Participants answered the following questions after every trial {inside the AR application}: "Which material is it?". The names of the ten materials included in Task 1 (\cref{sec:stimulus_t1}) constituted the answer options.
In addition, we wanted to assess their perception of the material's physical properties. We asked them to rate the density of the material - "How dense is the material?" - on a {continuous} scale from "As airy as milk foam" (0) to "As dense as gold" (100). They also rated the stiffness - "How stiff is the material?" - from "As elastic as rubber band" (0) to "As stiff as diamond" (100).
{These questions inside the AR application used visually uniform sliders without intermediate anchors, allowing participants to select any numeric value along a continuum.}

\vspace{0.8em}
\noindent\textbf{Confidence \& Helpfulness}
In a post-{condition, desktop-based} questionnaire, participants responded to three 7-point Likert scale questions on a scale from strongly disagree (1) to strongly agree (7) to assess their subjective confidence in their material and material properties answers {for all trials in the condition}. The items were: "I was confident in my material assignments.", "I was confident in my density estimations.", and "I was confident in my stiffness estimations.".
They furthermore answered the 7-point Likert scale question, "The audio feedback was helpful for classifying the materials." using the same scale.

\vspace{0.8em}
\noindent\textbf{Sound Realism}
To assess the sonic interaction realism, participants responded to the question "How realistic was the sonic interaction?" on a {continuous 0-100 slider} from not realistic at all (0) to absolutely realistic (100) {after} every material {inside the AR application.}

\subsection{Task 2}
In Task 2, participants were given two visually similar materials at the same time in each trial. The participants were again tasked to tap on the materials. Based on the visual and auditory cues, they had to identify the materials, rate their confidence in the assignment, and rate the helpfulness of the sound in distinguishing the materials.

\subsubsection{Stimulus}
\label{sec:stimulus_t2}
We purposefully selected six pairs of visually similar materials, which, however, differed in their actual materiality. In addition, we selected pairs to form two groups:

\vspace{0.8em}
\noindent\textbf{Ambiguous}  Material pairs that are visually and audibly similar: wood \& wood-printed plastic, glossy paper \& glossy plastic, stone \& stone-printed plastic.

\vspace{0.8em}
\noindent\textbf{Unambiguous}  Material pairs that are visually similar, but audibly different: glass \& plexiglass, rubber \& milky glass, coated ceramic \& coated wood.

\subsubsection{Independent Variables}

\noindent \textbf{{Sound}}
Task 2 again featured the two audio conditions: {material-based} and {generic}. In the {material-based} condition, the material segmentation model {was supplemented with marker tracking to ensure congruent sounds}, as visually indistinguishable materials {cannot} be reliably differentiated by {current} vision-based model{s}. Nevertheless, we included Task 2 to investigate whether material-based sonification could aid users in disambiguating visually similar materials. To ensure consistency, we reused the same material-based sonification model and physical parameters as in Task 1 for generating congruent sounds. 
In the {generic} condition, the same audio sample as in Task 1 was applied to all materials.

\vspace{0.8em}
\noindent \textbf{Audiovisual Ambiguity}
As described in \cref{sec:stimulus_t2}, we included two groups of material pairs, a group of visually and auditorily ambiguous materials and a group of visually ambiguous but auditorily distinct materials.

\subsubsection{Dependent Variables}

\noindent \textbf{Material} For each material pair, participants were asked to identify which material corresponded to which sample. They were not allowed to choose the same material for both samples. The items were called: "Which material is on the left?" and "Which material is on the right?". The answer options were, for example, "Glass" and "Plexiglass".

\vspace{0.8em}
\noindent \textbf{Confidence \& Helpfulness} 
Participants rated their confidence in the material assignments and the helpfulness of the sounds on a {continuous slider (0-100)} from Not confident/helpful at all (0) to Very confident/helpful (100) {after} every pair {inside the AR application}. The items were called "How confident are you in your material assignments?" and "How helpful was the sound to distinguish the two materials?".

\subsection{Participants}
24 participants (12 women, 12 men), with a mean age of 29.54 years (SD = 3.91), took part in the study. Most were PhD or master's students from the fields of biomedical engineering or computer science, with the rest consisting of professionals from very diverse subject areas ranging from business to literature studies and law. The subjects' music experience varied widely. One participant reported being a trained musician, four participants played instruments regularly, twelve of the participants knew how to play one or more instruments but rarely played, and seven participants reported not having any musical training. Experience with using AR or VR was similarly mixed. Fifteen subjects had used it either once or a few times before, eight reported regular or even daily use, and one had never used it before. Gaming habits were diverse as well, with seven people playing daily or weekly, but the rest reported to play very infrequently or never.

\subsection{Procedure}
Participants were seated in a quiet room at a well-lit table. After providing informed consent and completing a demographics survey, they were equipped with the AR headset, and eye calibration was performed. {The material samples were placed out of sight of the participants until retrieved one at a time only for the duration of the trial before being removed from sight again.} The audio was delivered to the participants via over-ear headphones.
Each task began with a training scene to familiarize participants with the procedure and AR environment. In Task 1, participants completed 10 trials in the first condition followed by a post-{condition} questionnaire, then proceeded to the second condition and again the questionnaire. 
In Task 2, they again started with a training scene followed by both conditions of each 3 trials. 
The order of conditions in both tasks was counterbalanced using Latin-square randomization.
The study concluded with a post-study questionnaire, where participants provided qualitative feedback.

\section{Results}
\label{sec:results}

\subsection{Task 1}
\begin{figure}[h!]
 \centering
 \includegraphics[width=\columnwidth]{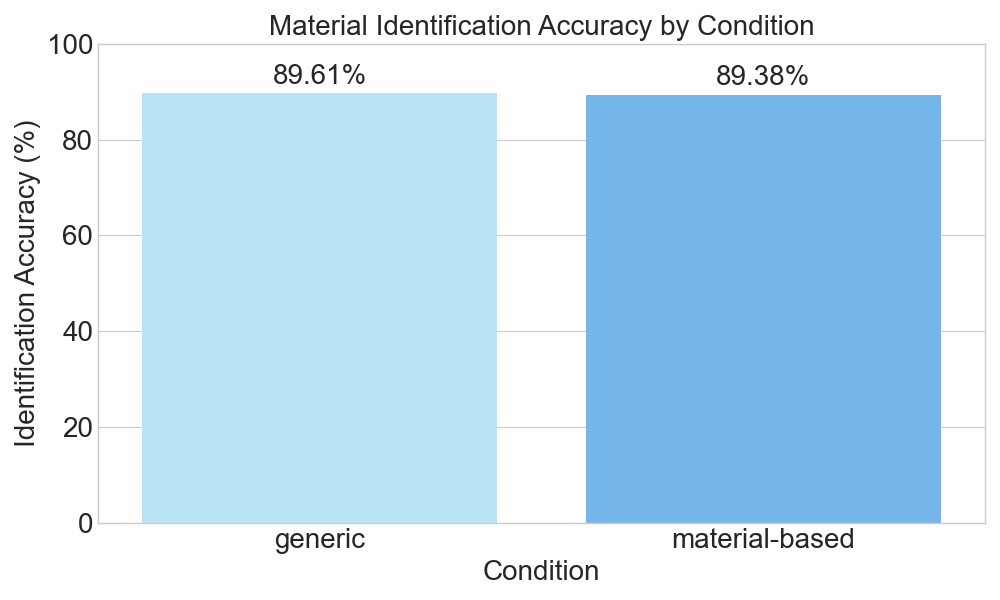}
 \caption{Material identification accuracy by condition in Task 1}
 \label{fig:t1_accuracy}
\end{figure}

A Shapiro-Wilk test showed a non-normal distribution of the data for all measures. Outliers in the time data were assessed using the interquartile range method, resulting in the removal of 5\% of the data points identified as outliers. No outliers were found for the other measures. Wilcoxon signed-rank tests showed significant differences (p $<$ 0.001) between the two sound conditions for all variables except the material recognition accuracy (see \cref{fig:t1_accuracy} and \cref{tab:means_std_t1}).

\subsubsection{{Material-based} vs. {Generic} Sounds}
\begin{figure*}[ht!]
    \centering
    \includegraphics[width=\textwidth]{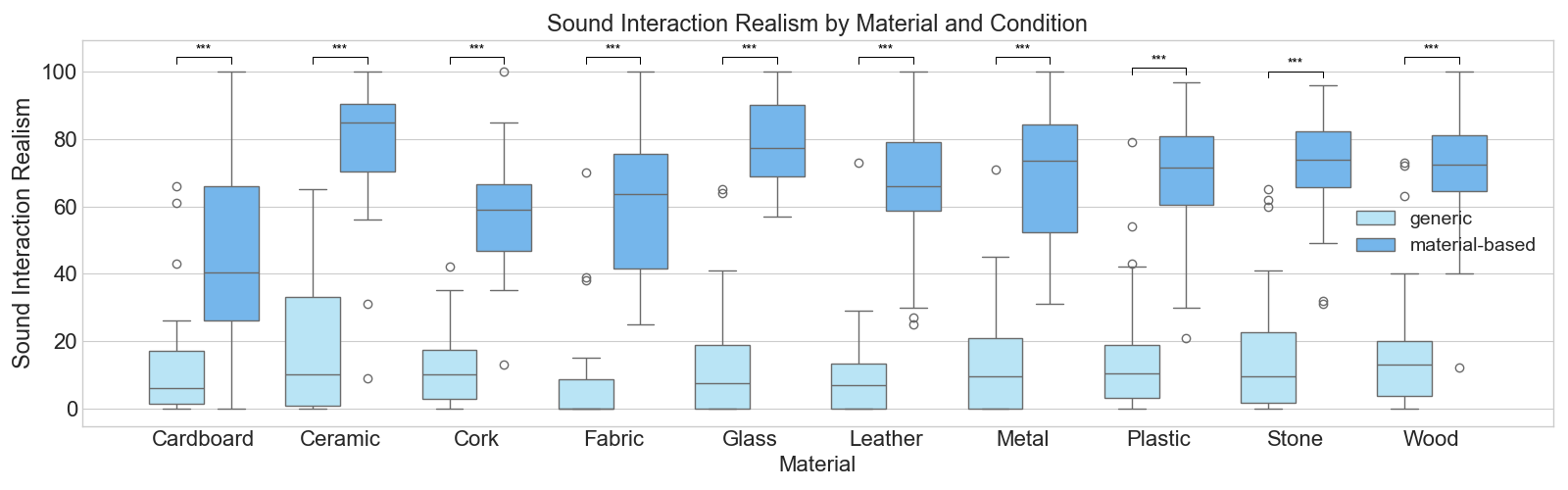}
    \caption{Sonic interaction realism ratings by material for the congruent and incongruent sounds in Task 1, *** = p $<$ 0.001}
    \label{fig:realism_t1}
\end{figure*}

There was a significant main effect of condition (p$<$0.05) on task time. Participants took significantly longer in the {material-based} condition ({46.83$\pm$16.02} seconds) than the {generic} condition ({41.44$\pm$16.01} seconds). The time per {trial} did not significantly vary based on which material they were viewing (p=0.4686).
There was also a significant main effect of condition (p$<$0.05) on the density and stiffness estimations. Materials were rated significantly more dense and more stiff in the {material-based} than in the {generic} condition.
The sonic interactions were rated significantly more realistic for the {material-based} condition (p$<$0.001). A per-material comparison of the realism rating for the materials used in the study can be found in \cref{fig:realism_t1}.

\begin{figure}[hb!]
 \centering
 \includegraphics[width=\columnwidth]{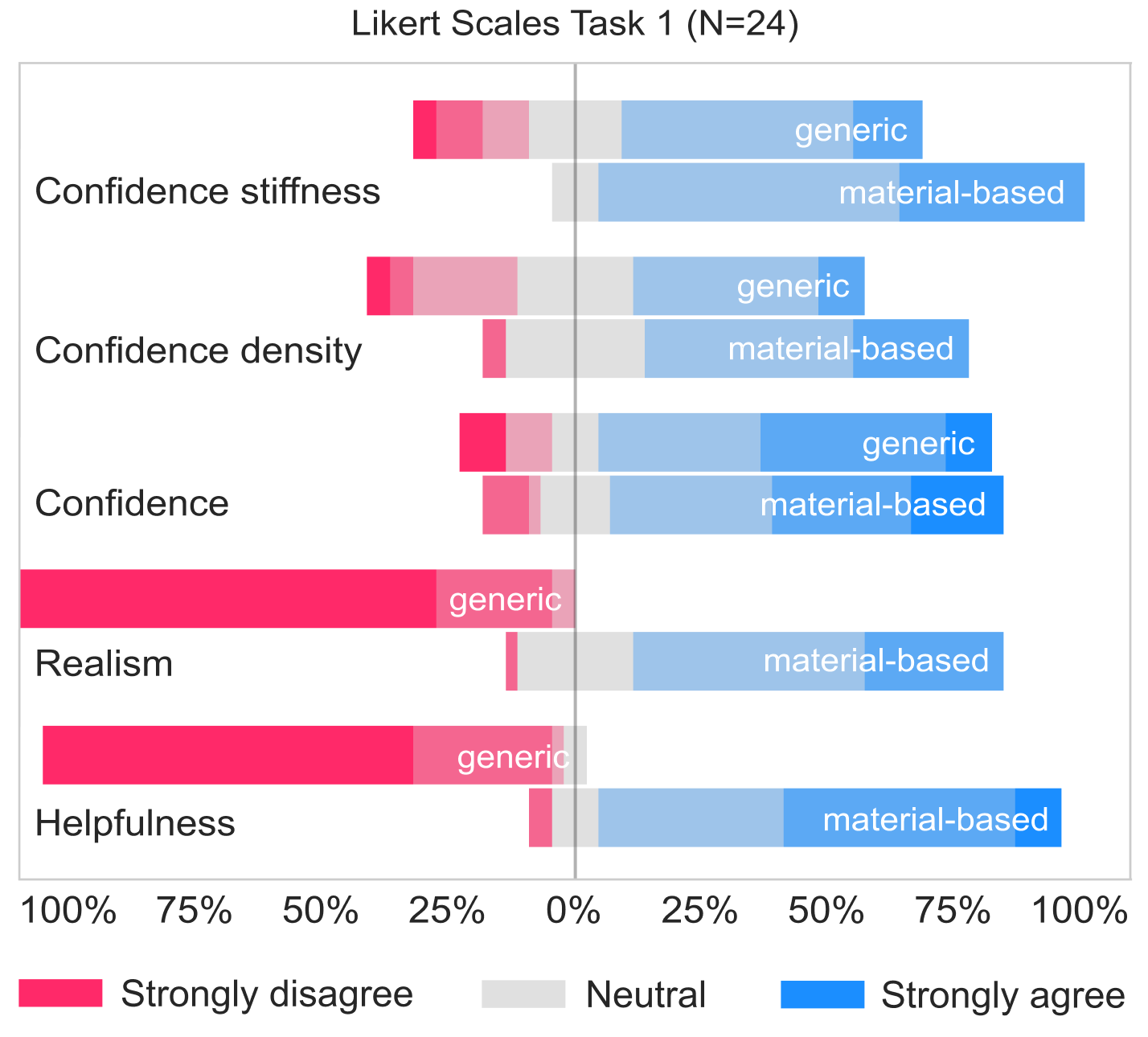}
 \caption{Barplot of Likert Scale responses in Task 1 by question and condition}
 \label{fig:likert_scale}
\end{figure}

Participants furthermore rated the {material-based} sounds ({Mdn=5.5, MAD=0.5}) much more helpful than the {generic} sound ({Mdn=1.0, MAD=0.0}) for material {identification} (p$<$0.001). There was no effect of condition on the subjects' material {identification} confidence. However, there was a significant effect of condition on material properties estimation confidence (density: p$<$0.05, stiffness: p$<$0.01) (see \cref{fig:likert_scale}).

\begin{table}[ht!]
  \caption{{Results} Task 1: {material identification accuracy} (percent); means, standard deviations and p-values of density (0-100), stiffness (0-100), time per trial (seconds){, sound realism (0-100); medians, median absolute deviations, p-values of the 7-point Likert scales on sound helpfulness, material confidence, density confidence, stiffness confidence}}
  \label{tab:means_std_t1}
  \scriptsize%
  \centering%
  \begin{tabu}{lccc}
  \toprule
Task 1 & {Generic} & {Material-based} & P-value \\
\midrule
Material Accuracy      & 89.61\%   & 89.38\%  & $>$ 0.05 \\
Density                & 54 $\pm$ 24   & 59 $\pm$ 23  & \textbf{$<$ 0.001} \\
Stiffness              & 52 $\pm$ 28  & 58 $\pm$ 27  & \textbf{$<$ 0.001} \\
Task Time              & 41.44 $\pm$ 16.01  & 46.83 $\pm$ 16.02  & \textbf{$<$ 0.001} \\
Sound Realism        & 15 $\pm$ 19      & 66 $\pm$ 21     & \textbf{$<$ 0.001} \\
Sound Helpfulness    &  { 1.0 (0.0)}  &  { 5.5 (0.5)} & \textbf{$<$ 0.001} \\
Material Confidence    & {5.0 (1.0)} & {5.0 (1.0)} & $>$ 0.05 \\
Density Confidence     & {4.0 (1.0)}  & {5.0 (1.0)} & \textbf{$<$ 0.05}\\
Stiffness Confidence  & { 5.0 (1.0)} & {5.0 (0.0)} & \textbf{$<$ 0.01} \\
  \bottomrule
  \end{tabu}%
\end{table}

\subsubsection{Hard vs. Soft Materials}
We further divided the materials into hard (Ceramic, Glass, Metal, Plastic, Stone, Wood) and soft (Leather, Cardboard, Cork, Fabric).
The sonic interaction was perceived to be significantly more realistic for hard materials (73$\pm$14) than soft materials (55$\pm$17) within the {generic} (p$<$0.01) and {material-based} (p$<$0.001) condition.
{Material-based} vs. {generic} sounds led to similar material {identification} accuracy for both soft and hard materials.

\subsection{Task 2}
The data showed a non-normal distribution. Outliers were removed from the task time data using the interquartile range method. Three outliers were identified. Please refer to \cref{tab:means_std_t2} for a breakdown of all the results in Task 2.

\subsubsection{{Material-based} vs. {Generic} Sounds}
\begin{figure*}[ht!]
    \centering
    \begin{minipage}{0.32\textwidth}
        \centering
        \includegraphics[width=\linewidth]{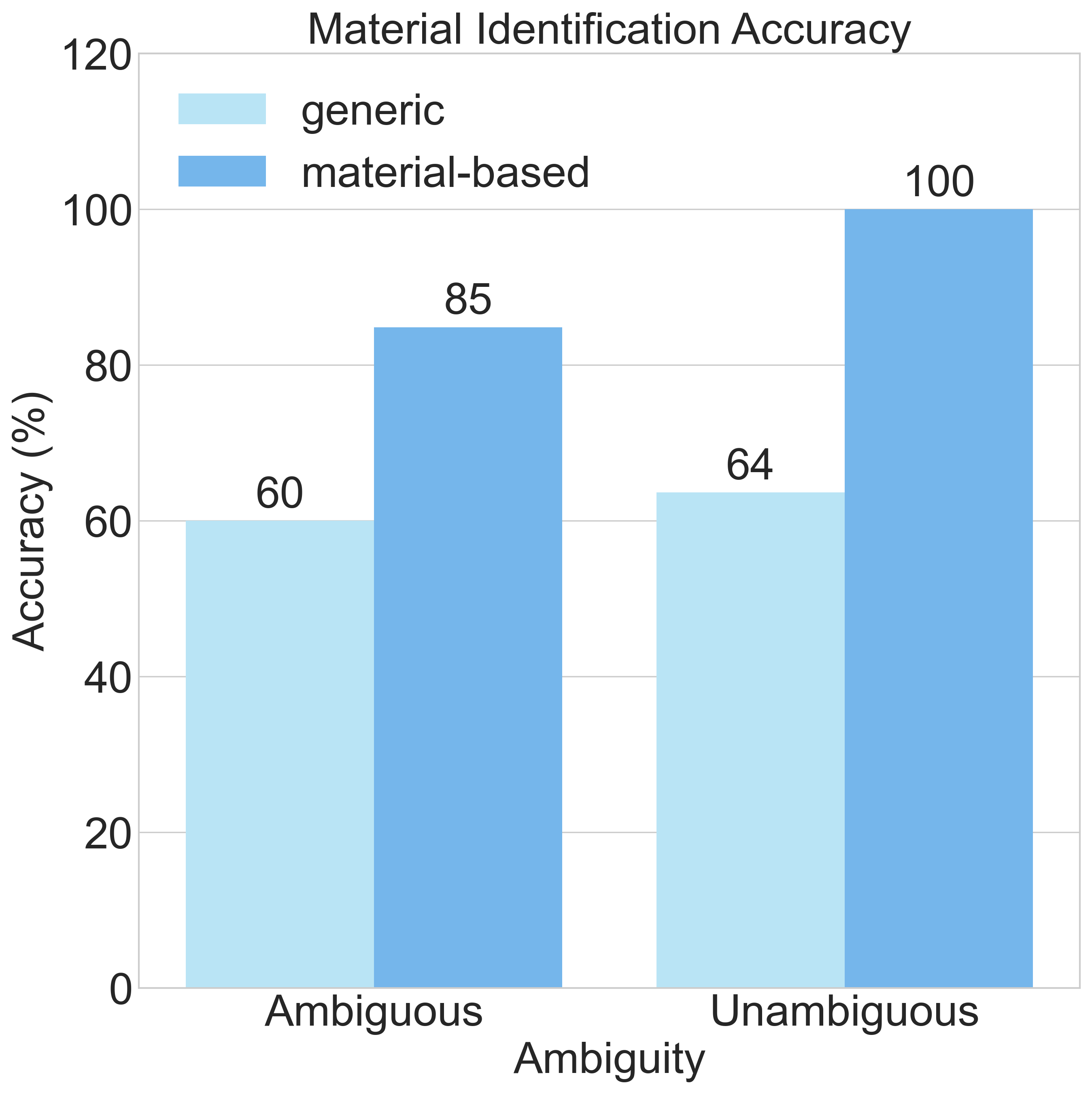}
        \label{fig:accuracy_t2}
    \end{minipage}%
    \hfill
    \begin{minipage}{0.32\textwidth}
        \centering
        \includegraphics[width=\linewidth]{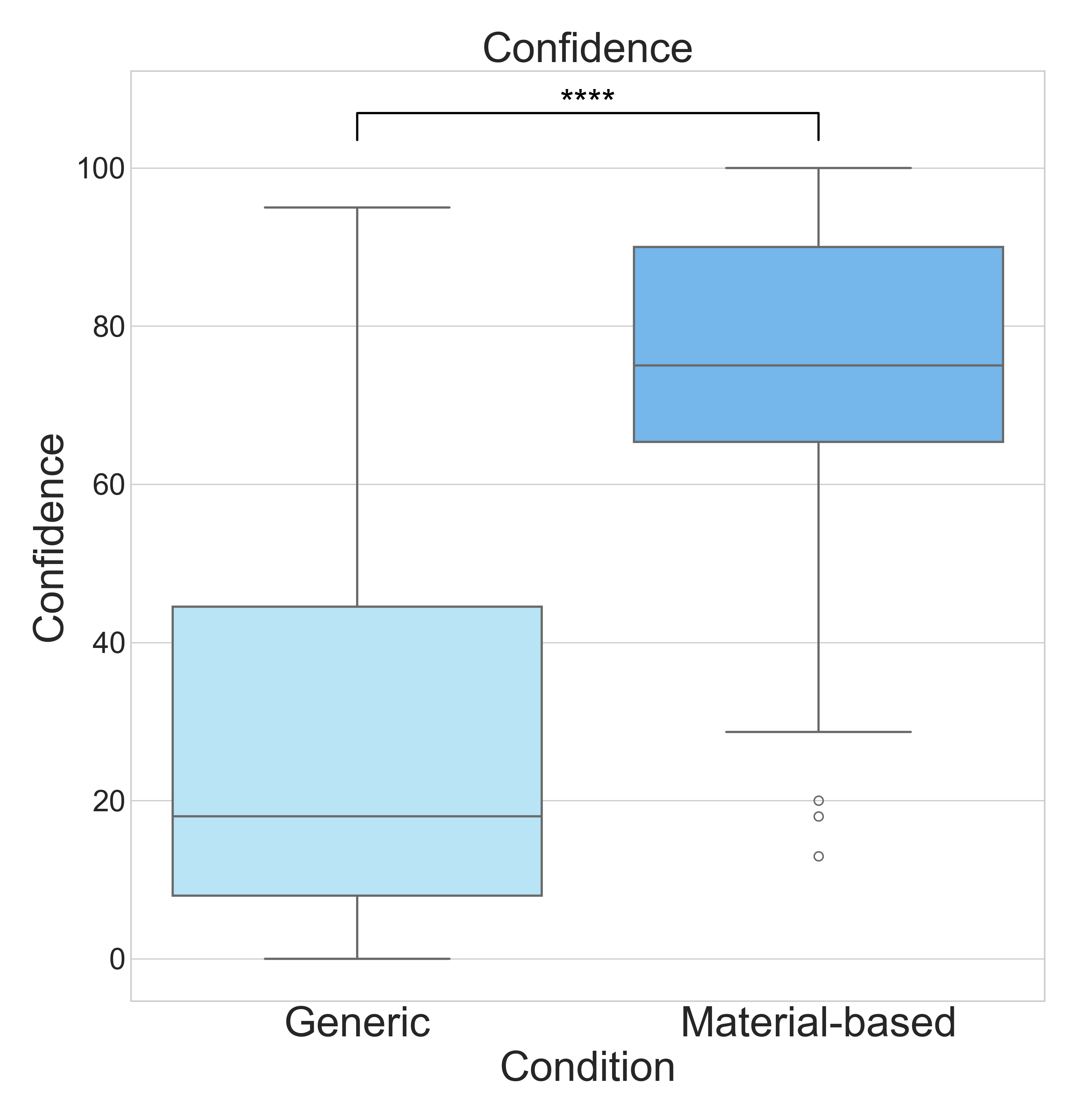}
        \label{fig:confidence_t2}
    \end{minipage}%
    \hfill
    \begin{minipage}{0.32\textwidth}
        \centering
        \includegraphics[width=\linewidth]{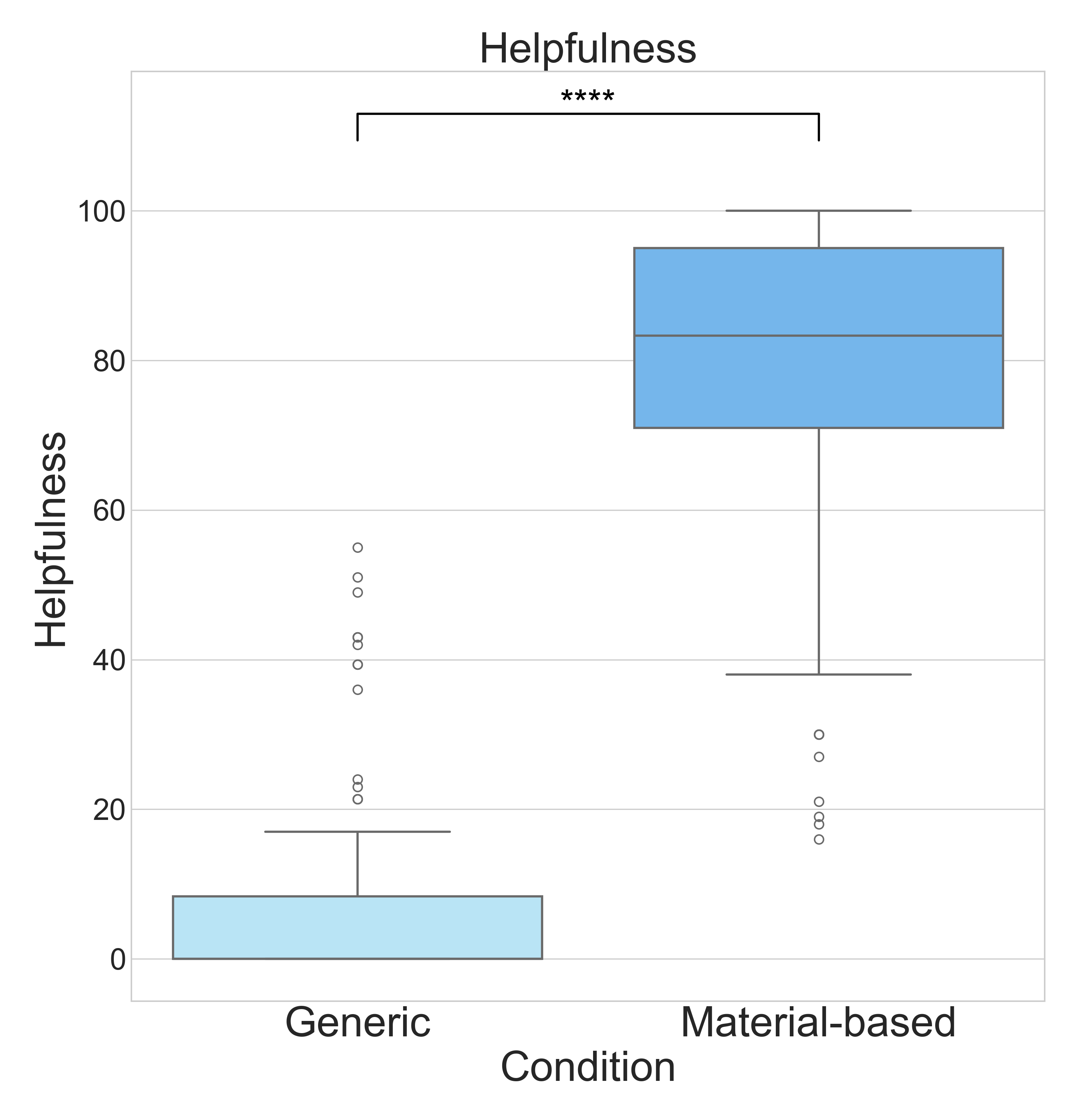}
        \label{fig:helpfulness_t2}
    \end{minipage}
    \caption{Task 2 results: Material identification accuracy by condition and auditory ambiguity (Ambiguous: Material pairs with visual and auditory similarity, Unambiguous: Material pairs with visual similarity, but audible differences), material assignment confidence by condition, sound helpfulness by condition}
    \label{fig:all_plots_t2}
\end{figure*}

There was a significant main effect of condition (p$<$0.05) on task time. Participants took significantly longer in the {material-based} (84.31$\pm$39.34) than {generic} (72.70$\pm$27.95) condition. 
The {material-based} sound feedback significantly improved participants' ability to correctly identify visually similar materials (Chi-square test: p $<$ 0.001), showing higher accuracy (92.75\%) than the {generic} condition (61.76\%).
This was also reflected in the confidence ratings, indicating significantly greater confidence in material assignments when using the congruent, {material-based} sounds (Wilcoxon signed-rank test: p$<$0.001) (see \cref{fig:confidence_t2}). 
Participants furthermore perceived the {material-based} sounds to be significantly more helpful in distinguishing between two materials (Wilcoxon signed-rank test: p$<$0.001) (see \cref{fig:helpfulness_t2}).

\subsubsection{Ambiguous vs. Unambiguous Material Sounds}

The analysis of differences \textbf{between} 
ambiguous (visually and auditorily similar) and unambiguous (visually similar yet auditorily distinct) material pairs showed no significant effect of auditory ambiguity on task time, sound helpfulness, or material identification accuracy. A significant main effect of auditory ambiguity on material identification confidence was found (p$<$0.001). Participants were significantly more confident in their material assignments for unambiguous (62.20\%) than ambiguous (40.21\%) pairs.

There was also a significant main effect of condition on material recognition accuracy \textbf{within} the ambiguous (p$<$0.05) and unambiguous (p$<$0.001) material groups. The {material-based} sound feedback resulted in significantly higher material recognition accuracy (84.80\%) than the {generic} sound feedback (60.00\%) within the auditorily ambiguous material pairs (see \cref{fig:accuracy_t2}). The same is true within the unambiguous material pairs, where {material-based} sounds (100\%) also achieved significantly greater material identification accuracy than {generic} sound feedback (63.60\%).

\begin{table}[tb]
  \caption{Means, standard deviations, and p-values of the measures in Task 2: material recognition accuracy (percent); material assignment confidence (0: Not confident at all, 100: Very confident), sound helpfulness (0: Not helpful at all, 100: Very helpful), time per trial (seconds)}
  \label{tab:means_std_t2}
  \scriptsize%
	\centering%
  \begin{tabu}{%
	l%
	*{3}{c}%
	}
  \toprule
   Task 2 & {Generic} & {Material-based} & P-value   \\
  \midrule
  Material Accuracy & 61.76\% &  92.75\% & \textbf{$<$ 0.001} \\
  Confidence & 28.22 $\pm$ 25.13 & 74.14 $\pm$ 19.20 & \textbf{$<$ 0.001} \\ 
  Helpfulness & 6.54 $\pm$ 12.32 & 79.16 $\pm$ 19.88 & \textbf{$<$ 0.001} \\
  Task Time & 72.70 $\pm$ 27.95 & 84.31 $\pm$ 39.34 & \textbf{$<$ 0.05} \\
  \bottomrule
  \end{tabu}%
\end{table}

\section{Discussion}
\label{sec:discussion}
{We introduced a framework for material-based sonic interactions in AR and demonstrated improved interaction realism over generic sounds, which are the current standard in AR applications.}

{More specifically, o}ur results showed that congruent audiovisual feedback in AR generated using our material-based approach leads to significantly more accurate material identification (p $<$ 0.001)) of real-world objects when distinguishing between two visually similar materials (H4). However, no significant difference in identification accuracy was observed in Task 1, where participants identified a single material at a time (H1). This is likely because most of the materials in Task 1 were already visually distinct enough to be reliably identified by sight without the need for matching impact sounds.
In everyday contexts, when conflicting visual and auditory cues are present, we often rely on additional haptic feedback to discern materials.
In our study, participants were not allowed to touch the materials, as audiovisual stimuli were the subject of examination.
It is also noteworthy that in real-world settings, the geometry of an object presents strong cues on object materiality. Through experience, we have learned to associate certain materials with certain object shapes. 
When visual information on object geometry and surface texture is combined, humans can infer material types more confidently than in our controlled study setting, where shape cues were missing. The use of uniformly shaped material plates might have led to material misclassifications in cases where {material-based} sounds were ambiguous.

The material identification results were consistent with participants' subjective confidence ratings (H2). In Task 1, where identification accuracy did not differ between conditions, confidence ratings were similarly unaffected. In contrast, in Task 2, where {material-based} sounds significantly improved identification accuracy, participants also reported significantly higher confidence in their choices (p $<$ 0.001) when audiovisual congruence was given.
Interestingly, regardless of identification accuracy, participants expressed greater confidence in their estimations of material properties (density, stiffness). This suggests that congruent, material-based sounds carry meaningful information about physical attributes, allowing users to perceive these properties with significantly increased confidence (p $<$ 0.05).
These findings imply that semantically congruent audio stimuli establish a deeper perceptual link between users and their physical surroundings, even when interacting with the real world through virtual objects.

Furthermore, we were able to show that sounds created using our material-based approach led to significantly greater (p $<$ 0.001) sound realism (H3).
The average sonic interaction realism was rated 66 out of 100. This result is in line with findings from a study in psychoacoustics that employed the same 0-100 scale for assessing subjective judgment of sound realism. They reported an average score of 68 for recordings of real material interaction sounds, compared to a mean score of 45 for the corresponding sound effects \cite{heller2022effects}. These findings demonstrate that even real sound recordings are not perceived as fully realistic, highlighting the need to interpret realism ratings relative to other sound conditions.
However, there is room for improvement in enhancing the realism of the {material-based} sounds. The sounds of materials with lower density and stiffness, like fabric, cardboard, cork, and leather, were rated as less realistic than rigid materials. This may be related to the combination of high damping coefficients and low stiffness in soft materials, resulting in lower frequency and shorter temporal evolution of the sounds, thereby offering participants less time to perceive differences in acoustic properties.
Alternative sound synthesis techniques, more suitable to soft bodies, e.g., data-driven approaches \cite{kun2023physicsdiff, su2018procedurally} could be explored to achieve greater sonic interaction realism for soft materials.

Lastly, we saw increased task time during the {material-based} condition in both tasks. In the {generic} condition, the same sound effect was played for every material. Therefore, a potential influence on faster trial time may be that the auditory memory only had to refer to one type of sound in every trial, speeding up the decision-making process.
Since the {material-based} sounds carry nuanced information about the materials' properties represented in sound parameters such as frequency and decay \cite{klatzky2000perception}, corresponding to each material, the sound varied greatly for every trial. This led participants to tap the materials more often in the {material-based} condition, carefully listening to the complex sound qualities unfold. This was especially true for Task 2, where the {material-based} trials consisted of two multifaceted sounds, while the {generic} trials only presented the same sound for both materials. As a result, participants had to integrate more information in the {material-based} condition to compare and judge the materials, prolonging the trial time.

\subsection{Limitations}

{One limitation of our study is the use of} average material property values for the material sounds. However, these values can largely impact the resulting sounds, making them less or more realistic. A potential solution could be two-fold: Firstly, establishing a material segmentation model that is able to provide more granular material classification output and secondly, using material parameter estimation methods, as proposed in Ren et al. \cite{ren2013example}, to estimate ideal parameters from the vision-based material output to achieve more realistic sounds.

Although both the real and the virtual object are involved in the sound-producing interaction, our study only sonified the tapping actions on the real surfaces in the scene, purposefully excluding the modeling of the virtual object (stick) involved in the collision to examine the material-based approach in an isolated manner.
However, to achieve a comprehensive simulation of the interaction, a bidirectional interaction between both objects, the real and the virtual, should be modeled. On top, adding more diverse action types like sliding, bowing, or scratching would further enhance the interaction capabilities.

In our study, we decided against including a condition without sound, as we were interested in the interplay of multiple senses. However, although the generic audio was incongruent, it still provided an auditory cue, making it difficult to assess its specific influence, e.g., on material identification accuracy{, sound helpfulness, and confidence}.
{
Generic sounds, beyond being non-helpful, can even be detrimental to the user’s ability to correctly identify the material of a real object, especially when the visual appearance of a material is ambiguous. For example, if an object could be either glass or plexiglass, and the generic sound is more similar in pitch and reverb to the impact sound of plexiglass, users may unconsciously rely on this misleading auditory cue, leading to incorrect identification. This can further lead to a misjudgment of decision confidence and sound helpfulness. 
In contrast, the absence of sound, i.e., in a purely visual condition, might limit confidence but does not introduce conflicting sensory information that biases the user’s decision. 
This perceptual asymmetry is important. While silence preserves ambiguity, generic sounds introduce false specificity. In this sense, generic sounds function not as neutral placeholders but as active confounders, potentially resulting in worse results than a silent condition.
Despite the confounders that a generic sound introduces, we purposely chose to use it in our baseline condition to mirror the current industry standard. 
Our goal was not to compare against silence, but to investigate whether material-based audio offers perceptual benefits over existing, generic audio design. 
However, we want to emphasize that our findings call attention to the potential negative perceptual consequences of using generic audio in current commercial XR systems and urge headset manufacturers and software developers to consider these modulatory perceptual effects when making audio design choices.}

\subsection{Application Areas}
Multiple potential application areas for the proposed sonification framework come to mind.
For one, the material-based sounds could be used to convey more detailed information about object materiality to blind or visually impaired users. 
Multiple works have shown that audio augmented reality can support blind or visually impaired users in perceiving and navigating their environment \cite{ribeiro2012auditory, doel2004geometric, kaul2021mobile}.
Our system could, for example, be used to augment a white cane with more granular sensory information about the materiality of the ground to facilitate navigation and support tactile understanding of the environment.

This material-based sonification method could furthermore be applied to medical use cases. Schütz et al. \cite{schuetz2024mmii} already introduced a physics-based sonification approach for medical applications such as tumor localization. While their work is limited to pre-defined models of the human anatomy, the framework presented in this paper could expand the use of physics-based sounds to medical AR applications. In minimally invasive laparoscopic procedures, for instance, the surgeon inserts instruments into the abdominal region via small incisions in the skin. This rids the surgeon of direct visual, audio, and haptic interaction with the anatomy. In this case, real-time audio feedback based on endoscopic images from inside the body could provide important textural information about the human tissue and thus increase awareness of human tissue properties. This could be helpful in distinguishing between cancerous and healthy tissue, potentially enhancing surgery outcomes.
However, as our study showed, sounds can influence or alter our perception of materiality. As sound has the power to influence our decision-making and behavior \cite{haas2022deceiving}, sonic interactions in surgical applications must be designed with careful consideration of their perceptual impact. To ensure high safety standards in medical applications, robust auditory cues should be targeted.

Besides the outlined use cases in accessibility and medical technology, material-based sonification could enhance AR training applications where users interact with virtual tools on real objects (e.g., assembling parts). Here, material-based audio could improve realism and skill transfer. In addition, telepresence in remote collaboration settings \cite{yu2021avatars} could benefit from realistic interaction sounds to make remote user actions feel more believable, thereby improving presence.

\subsection{Future Work}
A possible extension of our work could be the inclusion of the real object's shape, size, and context into the sonification pipeline. Achieving this within a modal synthesis framework would require real-time 3D reconstruction to obtain object geometries. With rapid advances in computer vision and the increasing computational capabilities of AR headsets, this may soon become feasible.

Touch interactions with physical objects are accompanied by visual, auditory, and haptic feedback. Touch was not considered in our work, but is instrumental to achieving fully realistic interaction experiences. Future work could investigate the addition of this sensory modality.

{It would also be interesting to investigate the impact of congruent and incongruent material sounds on user perception. This would necessitate the inclusion of a condition presenting ever-changing, incongruent material sounds instead of the constant generic sound used in our study.
We also did not include a "no sound" or a "real sound" condition. We see benefits in including either or both in future studies where appropriate.
A real sound condition could be especially relevant in work primarily focused on achieving high-fidelity sound synthesis. To implement a "real sound" condition, high-quality, controlled audio recordings of all relevant material-object interactions would be required. These recordings would then need to be mapped to specific interactions in the AR system, requiring precise real-time collision detection, timing synchronization, and spatial audio rendering to ensure perceptual alignment. Implementing such a control condition introduces significant technical complexity, but it might be worth incorporating in a study focused on audio fidelity.}

\section{Conclusion}
\label{sec:conclusion}
We introduced a material-based sonification framework that uses scene understanding and physical modeling sound synthesis to generate context-aware sound for AR interactions in real-time. The resulting material-based sounds establish audiovisual congruence during real-virtual object interactions in AR. 
In a user study comparing our congruent, material-based sounds to incongruent, generic sound effects, participants rated interactions with {material-based sounds} as significantly more realistic. Furthermore, {material-based} sounds enabled more accurate and confident differentiation of visually similar materials. 
These results suggest that context-aware, material-based auditory feedback can strengthen the perceptual connection between users and their physical environments during AR interactions. We hope that future research will build upon our findings and that commercial AR systems will increasingly incorporate realistic, context-aware sound feedback to {enhance} user experience.

\acknowledgments{
The authors wish to thank Prof. Dr. Stephan Krusche at the Technical University of Munich for his support with managing Apple Developer certificates.}

\bibliographystyle{abbrv-doi-hyperref-narrow}

\bibliography{sona}
\end{document}